\documentclass[pra,amsmath,amssymb,two column]{revtex4}
\usepackage{mathrsfs}
\usepackage{amssymb}

\input epsf.tex
\usepackage{graphicx}
\usepackage{amsthm}

\begin{document}

\title{Three-observer classical dimension witness violation with weak measurement}
\author{Hong-Wei Li$^{1,2,3,*}$, Yong-Sheng Zhang$^{2,3}$, Xue-Bi An$^{2,3}$, Zheng-Fu Han$^{2,3}$, Guang-Can Guo$^{2,3}$}

 \affiliation
 { $^1$ Henan Key Laboratory of Quantum Information and Cryptography,
Zhengzhou Information Science and Technology Institute, Henan, Zhengzhou 450000, China\\
 $^2$ Key Laboratory of Quantum Information, University of Science and Technology of China, Hefei, 230026, China \\
 $^3$ Synergetic Innovation Center of Quantum Information $\&$ Quantum Physics, University of Science and Technology of China, Hefei, 230026,
 China}

\begin{abstract}
Based on weak measurement technology, we propose the first
three-observer dimension witness protocol in a prepare-and-measure
setup. By applying the dimension witness inequality based on the
quantum random access code and the nonlinear determinant value, we
demonstrate that double classical dimension witness violation is
achievable if we choose appropriate weak measurement parameters.
Analysis of the results will shed new light on the interplay between
the multi-observer quantum dimension witness and the weak
measurement technology, which can also be applied in the generation
of semi-device-independent quantum random numbers and quantum key
distribution protocols.
\end{abstract}
\maketitle

{\it Introduction  - } Dimension is an important resource in quantum
information theory, for instance, a high dimensional quantum system
can enhance the performance of the quantum computation \cite{dim1},
quantum entanglement \cite{dim2}, quantum communication complexity
\cite{dim3} and others, and it can also reduce the security of
certain practical quantum key distribution systems \cite{qkd}. To
estimate the lower bound of the dimensions of a physical system,
quantum dimension witness has been proposed, which has important
applications in the semi-device-independent quantum key distribution
and quantum random number generation
\cite{li1,li2,brunner1,brunner2,experiment1,experiment2}. Until now,
it has been demonstrated that two-observer classical dimension
witness violation can be achieved with the Bell inequality test,
quantum random access code test, and determinant value test
respectively \cite{DW1,DW2,DW3}, but whether the multi-observer
classical dimension witness violation can be obtained or not is
still an open question.

Similar to a quantum dimension witness, quantum nonlocality also
plays a fundamental role in quantum information theory, which can be
used to guarantee the security of device-independent quantum
information protocols \cite{DI1,DI2}. In a two-observer system, two
observers can perform independent measurements on their subsystem to
test the Clauser-Horne-Shimony-Holt (CHSH) inequality \cite{CHSH},
the violation of which certifies quantum nonlocality. Recently, it
has been demonstrated that nonlocality sharing among three observers
can be established by applying weak measurement technology
\cite{CHSH1,CHSH2,CHSH3}, which demonstrates that subsequent
measurements can not be described by using classical probability
distributions.

Inspired by the work of sharing nonlocality with weak measurement
technology, three-observer classical dimension witness violation
will be analyzed in this paper, in which we analyze the dimension
witness inequality based on the quantum random access code and the
nonlinear determinant value. The analysis result indicate that
double classical dimension witness violations can be realized. More
 interestingly, we demonstrate local and global randomness generation
 in the three-observer protocol, and the analysis method can also be
 applied to future multi-observer quantum network studies.

{\it Weak measurement protocol  - } Weak measurement is a powerful
method to extract less information about a system with smaller
disturbance \cite{weak1}, which has proven to be useful for signal
amplification, state tomography, solving quantum paradoxes and
others \cite{weak2,weak3,weak4,weak5,weak6}. In this work, we use
the weak measurement definition given in Refs.
\cite{CHSH1,CHSH2,CHSH3}, and the corresponding analysis model is
given in Fig. 1.

\begin{figure}[!h]\center
\resizebox{9cm}{!}{
\includegraphics{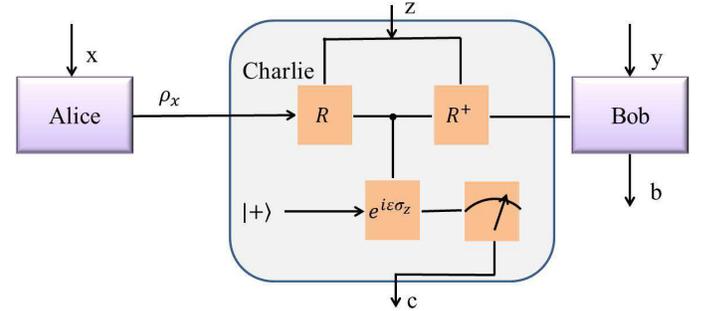}}
\caption{Analysis model of three-observer classical dimension
witness violation. Alice prepares the two-dimensional quantum state
$\rho_{x}$, and Bob and Charlie apply the two-dimensional strong
measurement and weak measurement respectively.}
\end{figure}

There are three observers Alice, Bob and Charlie in the analysis
model, and the purpose of our protocol is to establish double
classical dimension witness violations under the two-dimensional
Hilbert space restriction. More precisely, Alice prepares the
two-dimensional quantum state $\rho_{x}\in C^2$ and sends it through
the quantum channel with a different classical input random number
$x\in\{00,01,10,11\}$. Then, Charlie receives the quantum state and
applies the following operation with the input random number
$z\in\{0,1\}$
\begin{equation}
\begin{array}{lll}
R_z^+|0\rangle=|\omega_z\rangle,~~
R_z^+|1\rangle=|\omega_z^\perp\rangle,
\end{array}
\end{equation}
where the above operation illustrates that the rotation maps the
initial Hilbert space basis of $\{|0\rangle,|1\rangle\}$ to a new
basis of $\{|\omega_z\rangle,|\omega_z^\perp\rangle\}$ depending on
the given $z$ value. Charlie also has an two-dimensional ancillary
quantum state $|+\rangle=\frac{1}{\sqrt{2}}(|0\rangle+|1\rangle)$ in
the quantum channel, and then, the following control operation is
applied when Charlie receives the quantum state $|1\rangle$

\begin{equation}
\begin{array}{lll}
e^{i\epsilon\sigma_z}=\left(\begin{array}{ccc}
e^{i\epsilon}&0\\
0&e^{-i\epsilon} \end{array}\right).
\end{array}
\end{equation}
where $\epsilon$ is the weak measurement parameter. If Charlie
receives the quantum state $|0\rangle$, he will apply the identity
operation $I=\left(\begin{array}{ccc}
1&0\\
0&1 \end{array}\right)$, thus the total unitary operation in
Charlie's side can be given by

\begin{equation}
\begin{array}{lll}
U=R_z^+|0\rangle\langle0|R_z\otimes I +
R_z^+|1\rangle\langle1|R_z\otimes e^{i\epsilon\sigma_z}\\
~~~=|\omega_z\rangle\langle\omega_z|\otimes I +
|\omega_z^\perp\rangle\langle\omega_z^\perp|\otimes e^{i\epsilon\sigma_z}\\
 ~~~=W_B^{+z}\otimes I +W_B^{-z}\otimes e^{i\epsilon\sigma_z},
\end{array}
\end{equation}
where we define $W_B^{+z}=|\omega_z\rangle\langle\omega_z|$,
$W_B^{-z}=|\omega_z^\perp\rangle\langle\omega_z^\perp|$. Note that
Charlie will apply the identity operation $I$ if the weak
measurement parameter $\epsilon$ is $0$, which can be simply proved
since $W_B^{+z}\otimes I +W_B^{-z}\otimes I=I$. In this case,
Charlie will only obtain the initial ancillary quantum state
$|+\rangle$ after the control operation, which demonstrates that
Charlie can not gain any information about Alice's state $\rho_{x}$,
and there is no interaction introduced by Charlie correspondingly.

To prove the dimension witness in the three-observer system, we
should analyze the density matrix to illustrate Bob and Charlie's
system. The initial quantum state can be given by

\begin{equation}
\begin{array}{lll}
\rho_{BCx}=\rho_{x}\otimes |+\rangle\langle+|.
\end{array}
\end{equation}
By applying the previous unitary transformation $U$, the quantum
state $\rho_{BCx}$ will be transformed into

\begin{equation}
\begin{array}{lll}
\rho_{BCx}'=U\rho_{BCx}U^\dag\\
~~~~~~~=W_B^{+z}\otimes I \rho_{BCx}W_B^{+z}\otimes I\\
~~~~~~~~+W_B^{+z}\otimes I \rho_{BCx}W_B^{-z}\otimes e^{-i\epsilon\sigma_z}\\
~~~~~~~~+W_B^{-z}\otimes e^{i\epsilon\sigma_z}\rho_{BCx}W_B^{+z}\otimes I\\
~~~~~~~~+W_B^{-z}\otimes
e^{i\epsilon\sigma_z}\rho_{BCx}W_B^{-z}\otimes
e^{-i\epsilon\sigma_z}.
\end{array}
\end{equation}

The quantum state $\rho_{Bx}'=\text{Tr}_C\rho_{BCx}'$ will be
transmitted to Bob, as follows

\begin{equation}
\begin{array}{lll}
\rho_{Bx}'=\text{Tr}_C\rho_{BCx}'\\
~~~~~=W_B^{+z} \rho_x W_B^{+z}+W_B^{-z} \rho_x
W_B^{-z}\\~~~~~~~+\cos\epsilon (W_B^{+z} \rho_x W_B^{-z}+ W_B^{-z}
\rho_x
W_B^{+z})\\
~~~~~=(1-\cos(\epsilon))(W_B^{+z} \rho_x W_B^{+z}+W_B^{-z} \rho_x
W_B^{-z})\\
~~~~~~~+\cos(\epsilon) \rho_x .
\end{array}
\end{equation}
Similarly, the quantum state $\rho_{Cx}'=\text{Tr}_B\rho_{BCx}'$
will be transmitted to Charlie, as follows
\begin{equation}
\begin{array}{lll}
\rho_{Cx}'=\text{Tr}_B\rho_{BCx}'\\
~~~~~=\text{Tr}_B(W_B^{+z}\rho_x)|+\rangle\langle+|\\
~~~~~~~+\text{Tr}_B(W_B^{-z}\rho_x)e^{i\epsilon\sigma_z}|+\rangle\langle+|e^{-i\epsilon\sigma_z}.
\end{array}
\end{equation}

After receiving the quantum state $\rho_{Bx}'$, Bob will apply the
two-dimensional projective measurement depending on the input random
number $y\in\{0,1\}$. If Bob's input random number is $0$, the
corresponding measurement basis in Bob's side is given by

\begin{equation}
\begin{array}{lll}
\{|\nu_0\rangle\langle\nu_0|,
|\nu_0^\bot\rangle\langle\nu_0^\bot|\}.
\end{array}
\end{equation}
If Bob's input is $1$, Bob's measurement basis is
\begin{equation}
\begin{array}{lll}
\{|\nu_1\rangle\langle\nu_1|,
|\nu_1^\bot\rangle\langle\nu_1^\bot|\},
\end{array}
\end{equation}
where the measurement outcomes $|\nu_0\rangle\langle\nu_0|$ and
$|\nu_1\rangle\langle\nu_1|$ indicate classical bit $1$, and the
measurement outcomes $|\nu_0^\bot\rangle\langle\nu_0^\bot|$ and
$|\nu_1^\bot\rangle\langle\nu_1^\bot|$ indicate classical bit $-1$.

After receiving the quantum state $\rho_{Cx}'$, Charlie will apply
the two-dimensional projective measurement
\begin{equation}
\begin{array}{lll}
\{|t\rangle\langle t|, |t^\bot\rangle\langle t^\bot|\},
\end{array}
\end{equation}
where the measurement outcomes $|t\rangle\langle t|$ and
$|t^\bot\rangle\langle t^\bot|$ respectively indicate the classical
bit $1$ and $-1$. Before the state measurement, Charlie randomly
chooses rotation $\{R_z,R_z^+\}$ with respect to the input random
number $z\in\{0,1\}$.

In the case where the weak measurement parameter $\epsilon$ is $0$,
there is no interaction on Charlie's side, and then, Charlie's
density matrix will be transformed into $\rho_{Cx}'
=\text{Tr}_B(W_B^{+z}\rho_x+W_B^{-z}\rho_x)|+\rangle\langle+|=|+\rangle\langle+|$,
and Bob's density matrix will be transformed into $\rho_{Bx}' =
\rho_x $. In the following section, we will focus on the situation
$0\leq\epsilon\leq \pi$, and analyze the interaction introduced by
the weak measurement, which can be used to obtain the information
gain in Charlie's side.

The general representation of a qubit can be illustrated by using
the density matrix formalism $\frac{I+\vec{r}\cdot\vec{\sigma}}{2}$,
where $\vec{\sigma}$ is the Pauli matrix vector
$(\sigma_x=\left(\begin{array}{ccc}
0&1\\
1&0
\end{array}\right),\sigma_y=\left(\begin{array}{ccc}
0&-i\\
i&0 \end{array}\right),\sigma_z=\left(\begin{array}{ccc}
1&0\\
0&-1 \end{array}\right))$, and $\vec{r}=(r_x,r_y,r_z)$ is a real
three-dimensional vector such that $\|r\|\leq1$, thus the
conditional probability distribution to illustrate Alice and Bob's
system can be given by

\begin{equation}
\begin{array}{lll}
p(b|xy) =\sum\limits_{z}p(z)\text{Tr}(B_y^b\rho_{Bx}'(z))\\
~~~~~~~~~=\frac{1}{2}(1-\cos(\epsilon))\sum\limits_{z}p(z)b\omega_z\cdot\nu_y\times
 \omega_z\cdot\rho_x\\
~~~~~~~~~~~+\frac{1}{2}\cos(\epsilon)(1+b\nu_y\cdot\rho_x)+\frac{1}{2}(1-\cos(\epsilon)),
\end{array}
\end{equation}
where $B_y^b$ is the measurement operator acting on the
two-dimensional Hilbert space with input parameter $y$ and output
parameter $b$ by considering the prepared quantum state
$\rho_{Bx}'(z)$. The conditional probability distribution to
illustrate Alice and Charlie's system is given by

\begin{equation}
\begin{array}{lll}
p(c|xz)=
\text{Tr} (C^c\rho_{Cx}'(z))\\
~~~~~~~~~=\frac{1}{2}(1+\omega_z\cdot\rho_{x})(1+c-c\sin^2(\epsilon)),
\end{array}
\end{equation}
where $C^c$ is the measurement operator acting on the
two-dimensional Hilbert space to obtain the measurement outcome $c$
with the state $\rho_{Cx}'(z)$.

Based on the previous observed statistics $p(b|xy)$ and $p(c|xz)$,
we can calculate the dimension witness value between Alice and Bob
and between Alice and Charlie respectively.

{\it Quantum dimension witness based on quantum random access code -
}
 The first quantum dimension witness inequality based on the quantum
random access code in the two-observer system is given by
\cite{DW3,li1,li2}

\begin{equation}
\begin{array}{lll}
W_1=p(1|000)+p(1|001)+p(1|010)-p(1|011)\\
~~~~~-p(1|100)+p(1|101)-p(1|110)-p(1|111).
\end{array}
\end{equation}
In the two-dimensional Hilbert space, it has been proved that the
upper bound of the classical dimension witness value $W_1$ is $2$,
while the upper bound of the quantum dimension witness value $W_1$
is $2\sqrt{2}$. We will apply this dimension witness equation to
analyze the dimension witness values between Alice and Bob $W_{1AB}$
and between Alice and Charlie $W_{1AC}$.

Based on the state preparation and measurement model given in the
Appendix \cite{supp1}, the dimension witness value between Alice and
Bob is given by

\begin{equation}
\begin{array}{lll}
W_{1AB}=\sqrt{2}(\cos(\epsilon)+1).
\end{array}
\end{equation}
The dimension witness value between Alice and Charlie is given by

\begin{equation}
\begin{array}{lll}
W_{1AC}=2\sqrt{2}\sin^2(\epsilon).
\end{array}
\end{equation}
Note that $\epsilon=0$ indicates $W_{1AB}=2\sqrt{2}$ and
$W_{1AC}=0$, which demonstrates that Charlie's quantum state has no
interaction with Bob's system. In this case, only $W_{1AB}$ violates
the classical upper bound, which is reduced to a two-observer system
(Alice-Bob). Similarly, $\epsilon=\frac{\pi}{2}$ indicates
$W_{1AB}=0$ and $W_{1AC}=2\sqrt{2}$, which is reduced to a
two-observer system (Alice-Charlie). Thus, our protocol is more
general compared with the previous 2-observer protocols, which sheds
new light on the dimension witness in the network environment.

The detailed quantum dimension witness values $W_{1AB}$ and
$W_{1AC}$ with different weak measurement parameter $\epsilon$
values are given in Fig. 2.
\begin{figure}[!h]\center
\resizebox{9cm}{!}{
\includegraphics{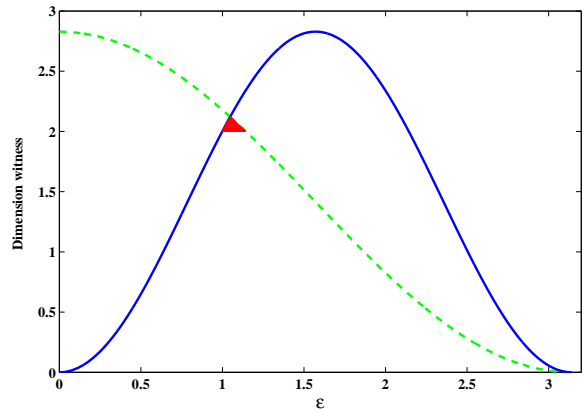}}
\caption{Three-observer classical dimension witness violation based
on the quantum random access code test. The dashed green line
corresponds to the quantum dimension witness value $W_{1AB}$, solid
blue line corresponds to the quantum dimension witness value
$W_{1AC}$, and red area corresponds to the double classical
dimension witness violation. }
\end{figure}
The analysis results indicate that double classical dimension
witness violations ($min\{W_{1AB},W_{1AC}\}>2$) can be obtained if
the weak measurement parameter $\epsilon$ satisfies
$\arcsin(\sqrt[4]{\frac{1}{2}})<\epsilon<\arccos(\sqrt{2}-1)$.

{\it Quantum dimension witness based on the determinant value - }
The second quantum dimension witness inequality based on the
nonlinear determinant value test in the two-observer system is given
by \cite{brunner2,experiment1}
\begin{equation}
\begin{array}{lll}
W_2=\left|\begin{array}{ccc}
p(1|000)-p(1|010)&p(1|100)-p(1|110)\\
p(1|001)-p(1|011)&p(1|101)-p(1|111)\end{array}\right|.
\end{array}
\end{equation}
Assuming that the state preparation and measurement devices are
independent, it has been proved that this nonlinear dimension
witness can tolerate an arbitrarily low detection efficiency. In the
two-dimensional Hilbert space, the upper bound of the quantum
dimension witness value is $1$, while the classical dimension
witness value is $0$. Similar to the previous subsection, the
dimension witness values between Alice and Bob $W_{2AB}$ and between
Alice and Charlie $W_{2AC}$ can be analyzed.

Based on the state preparation and measurement model given in the
Appendix \cite{supp2}, the dimension witness value between Alice and
Bob is given by

\begin{equation}
\begin{array}{lll}
W_{2AB}=(\frac{1}{2}\cos(\epsilon)+\frac{1}{2})^2.
\end{array}
\end{equation}
The dimension witness value between Alice and Charlie is given by

\begin{equation}
\begin{array}{lll}
W_{2AC}=\sin^4(\epsilon).
\end{array}
\end{equation}
Note that $\epsilon=0$ indicates $W_{2AB}=1$ and $W_{2AC}=0$, which
demonstrates that Charlie's quantum state has no interaction with
Bob's system. In this case, only $W_{2AB}$ violates the classical
upper bound, which is reduced to a two-observer system (Alice-Bob).

The detailed quantum dimension witness values $W_{2AB}$ and
$W_{2AC}$ with different weak measurement parameters $\epsilon$ are
given in Fig. 3.
\begin{figure}[!h]\center
\resizebox{9cm}{!}{
\includegraphics{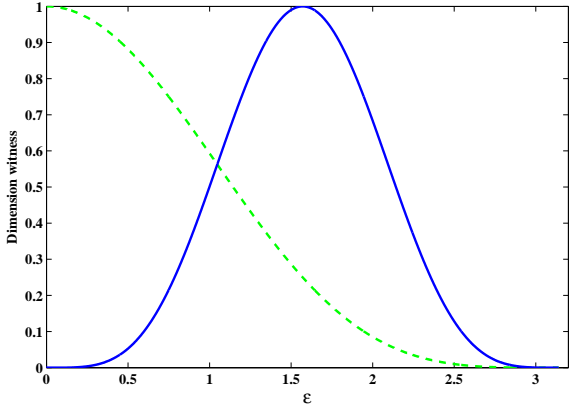}}
\caption{Three-observer classical dimension witness violation based
on the determinant value. The dashed green line corresponds to the
quantum dimension witness value $W_{2AB}$, and solid blue line
corresponds to the quantum dimension witness value $W_{2AC}$.}
\end{figure}
The analysis result demonstrates that double classical dimension
witness violations ($W_{2AB}>0, W_{2AC}>0$) can be obtained if
$0<\epsilon<\pi$. However, since Bob's system may be influenced by
Charlie's system, the classical dimension witness violation
$W_{2AB}>0$ can not be directly applied to guarantee the security of
the measurement outcome $b$.

{\it Semi-device-independent random number generator - } The
classical dimension witness violation can generate
semi-device-independent quantum random numbers, for which we can
only assume knowledge of the dimension of the underlying physical
system, but otherwise nothing about the quantum devices. The
generated random numbers in our protocol are the measurement
outcomes $b$ and $c$, and the eavesdropper can not guess the
measurement outcomes even if the state preparation and measurement
devices are imperfect.

Randomness generation can be divided into global randomness and the
local randomness, where global randomness must analyze the global
conditional probability distribution $p(b,c|x,y,z)$, while local
randomness must analyze the local conditional probability
distribution $p(b|x,y)$. With the given conditional probability
distributions, the random number generation efficiency can be
estimated by following min-entropy functions \cite{min}

\begin{equation}
\begin{array}{lll}
H_{min1}=-\log_2[\frac{1}{16}\sum\limits_{x,y,z}
\text{max}_{b,c}(p(b,c|x,y,z))],\\
H_{min2}=-\log_2[\frac{1}{8}\sum\limits_{x,y}
\text{max}_{b}(p(b|x,y))].
\end{array}
\end{equation}
To analyze the global randomness generation efficiency $H_{min1}$,
the maximal guessing probability $\frac{1}{16}\sum\limits_{x,y,z}
max_{b,c}(p(b,c|x,y,z))$ can be estimated by
\begin{equation}
\begin{array}{lll}
\frac{1}{16}\sum\limits_{x,y,z}\text{max}_{b,c}p(b,c|x,y,z)
\\
=\frac{1}{16}\sum\limits_{x,y,z}\text{max}_{b,c}p(b|x,y,z)p(c|x,y,z,b)\\
=\frac{1}{16}\sum\limits_{x,y,z}\text{max}_{b,c}(p(b|x,y,z)p(c|x,z))\\
\leq[\frac{1}{8}\sum\limits_{x,z}\text{max}_{c}p(c|x,z)]\times \text{max}_{b,x,y,z}(p(b|x,y,z),\\
\end{array}
\end{equation}
where the second line is based on the condition for which Charlie's
measurement outcome $c$ can not be effected by Bob, and the
corresponding randomness generation efficiency $H_{min1}$ is given
by

\begin{equation}
\begin{array}{lll}
H_{min1}\geq-\log_2[\frac{1}{8}\sum\limits_{x,z}\text{max}_{c}p(c|x,z))]\\
~~~~~~~~~~~~-\log_2[\text{max}_{b,x,y,z}(p(b|x,y,z)],
\end{array}
\end{equation}
where the first part is the randomness generation in Charlie's side,
and the second part is the randomness generation in Bob's side.

In the two-observer system, with a given random access code based
dimension witness value $W_1$, the relationship between $W_1$ and
the randomness generation efficiency $H'_{min2}(W_1)$ \cite{li3} is
given by
\begin{equation}
\begin{array}{lll}
H'_{min2}(W_1)=-\log_2(\frac{1}{2}+\frac{1}{2}\sqrt{\frac{1+\sqrt{1-(\frac{W_1^2-4}{4})^2}}{2}}).
\end{array}
\end{equation}
However, the eavesdropper can apply Charlie's input parameter $z$ to
guess Bob's measurement outcome $b$, thus the previous method can
not be directly applied in our protocol. To estimate the local
randomness generation in Bob's side, we will estimate the dimension
witness value between Alice and Bob $W_{1AB(z)}$ with different
input random number $z\in\{0,1\}$ as follows

\begin{equation}
\begin{array}{lll}
W_{1AB(z=0)}=W_{1AB(z=1)}=\sqrt{2}\cos(\epsilon)+\sqrt{2},
\end{array}
\end{equation}
with the detailed calculation is given in the Appendix \cite{supp3},
and the corresponding local randomness generation efficiency in
Bob's side is $H'_{min2}(W_1=\sqrt{2}\cos(\epsilon)+\sqrt{2})$.

 With the given determinate value based on dimension witness
$W_2$ in the two-observer system, the relationship between the
quantum dimension witness value $W_2$ and the randomness generation
efficiency $H''_{min2}(W_2)$ \cite{brunner2} is given by

\begin{equation}
\begin{array}{lll}
H''_{min2}(W_2)=-\log_2(\frac{1}{2}+\frac{1}{2}\sqrt{\frac{1+\sqrt{1-W_2^2}}{2}}).
\end{array}
\end{equation}
Similar to the previous calculation, the dimension witness value
between Alice and Bob $W_{2AB(z)}$ with different input random
number $z$ can be given by

\begin{equation}
\begin{array}{lll}
W_{2AB(z=0)}=W_{2AB(z=1)}=\cos(\epsilon),
\end{array}
\end{equation}
with the detailed calculation is given in the Appendix \cite{supp4},
and the corresponding local randomness generation efficiency in
Bob's side is $H''_{min2}(W_{2}=\cos(\epsilon))$.

Based on the previous analysis result, the detailed local randomness
generation efficiency with different weak measurement parameter
$\epsilon$ values are given in Fig. 4.
\begin{figure}[!h]\center
\resizebox{9cm}{!}{
\includegraphics{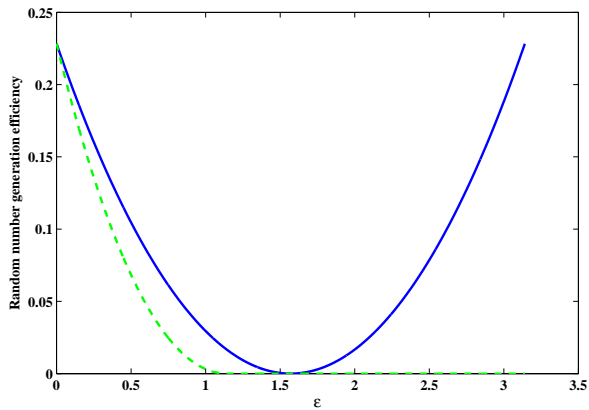}}
\caption{Local random number generation efficiency with different
weak measurement parameters $\epsilon$. The dashed green line
corresponds to $H'_{min2}(W_1=\sqrt{2}\cos(\epsilon)+\sqrt{2})$, and
solid blue line corresponds to $H''_{min2}(W_{2}=\cos(\epsilon))$. }
\end{figure}
Note that the local random number generation in Charlie's side can
be directly estimated by the two-observer protocol. Thus double
local randomness generation can be realized if we choose an
appropriate weak measurement parameter.

{\it Conclusion and Discussion  - } In conclusion, we proposed a
three-observer dimension witness protocol, where the weak
measurement technology was applied to analyze the double classical
dimension witness violations. The results of our analysis shed new
light on understanding the quantum dimension witness in the network
environment. The three-observer dimension witness protocol can be
assumed to be a sequential measurement protocol, and it will be
interesting to analyze a higher dimensional multi-observer quantum
system with sequential measurement \cite{sequential} technology. We
demonstrated the randomness generation in the three-observer
protocol, and the analysis results demonstrate that weak measurement
may have significant applications in multi-observer
semi-device-independent quantum information theory. This study also
provides tremendous motivation for further experimental research.

{\it Acknowledgements  - } We thank Marcin Paw{\l}owski for helpful
discussions. This work are supported by the National Natural Science
Foundation of China (Grant Nos. 61675235, 11304397, 11674306),
National key research and development program of China (Grant No.
2016YFA0302600) and China Postdoctoral Science Foundation (Grant No.
2013M540514). $^*$ To whom correspondence should be addressed,
Email: lihow@ustc.edu.cn.

\textbf{Author Contributions - } Hong-Wei Li, Yong-Sheng Zhang and
Guang-Can Guo conceived the project. Hong-Wei Li, Xue-Bi An and
Zheng-Fu Han performed the optimization calculation and analysis.
Hong-Wei Li wrote the paper.

 \textbf{Competing financial interests - }  The authors declare no
competing financial interests.

\section{ Appendices }

\subsection{ Quantum dimension witness based on quantum random
access code }
 The general representation of a qubit can be
illustrated by using the density matrix formalism
$\frac{I+\vec{r}\cdot\vec{\sigma}}{2}$, where $\vec{\sigma}$ is the
Pauli matrix vector $(\sigma_x=\left(\begin{array}{ccc}
0&1\\
1&0
\end{array}\right),\sigma_y=\left(\begin{array}{ccc}
0&-i\\
i&0 \end{array}\right),\sigma_z=\left(\begin{array}{ccc}
1&0\\
0&-1 \end{array}\right))$, and $\vec{r}=(r_x,r_y,r_z)$ is a real
three-dimensional vector such that $\|r\|\leq1$. In the following
section, we apply $\vec{r}=(r_x,r_y,r_z)$ to demonstrate the quantum
state $\rho_{x}$ $(x\in\{00,01,10,11\})$,
$|\nu_y\rangle\langle\nu_y|$ $(y\in\{0,1\})$,
$|\omega_z\rangle\langle\omega_z|$ $(z\in\{0,1\})$ and
$|t\rangle\langle t|$.

To prove the double classical dimension witness violation under Eq.
(13), we propose the following density matrix prepared in Alice's
side

\begin{equation}
\begin{array}{lll}
\rho_{00}=(\frac{1}{\sqrt{2}},0,\frac{1}{\sqrt{2}}),\\
\rho_{01}=(-\frac{1}{\sqrt{2}},0,\frac{1}{\sqrt{2}}),\\
\rho_{10}=(\frac{1}{\sqrt{2}},0,-\frac{1}{\sqrt{2}}),\\
\rho_{11}=(-\frac{1}{\sqrt{2}},0,-\frac{1}{\sqrt{2}}).
\end{array}
\end{equation}
The measurement operator in Bob's side is given by
\begin{equation}
\begin{array}{lll}
\nu_{0}=(0,0,1),\\
\nu_{1}=(1,0,0).
\end{array}
\end{equation}
The measurement operator in Charlie's side is given by
\begin{equation}
\begin{array}{lll}
t=(1,0,0).
\end{array}
\end{equation}
The rotation operator in Charlie's side is given by
\begin{equation}
\begin{array}{lll}
\omega_{0}=(0,0,1),\\
\omega_{1}=(1,0,0).
\end{array}
\end{equation}

Based on the previous state preparation and measurement operator, we
can calculate the corresponding conditional probability distribution
$p(b=1|x,y)_{AB}$ with different classical input random numbers
$x\in\{00,01,10,11\}$ and $y\in\{0,1\}$.

\begin{equation}
\begin{array}{lll}
p(1|000)_{AB}=\frac{\sqrt{2}}{8}\cos(\epsilon)+\frac{4+\sqrt{2}}{8},\\
p(1|001)_{AB}=\frac{\sqrt{2}}{8}\cos(\epsilon)+\frac{4+\sqrt{2}}{8},\\
p(1|010)_{AB}=\frac{\sqrt{2}}{8}\cos(\epsilon)+\frac{4+\sqrt{2}}{8},\\
p(1|011)_{AB}=-\frac{\sqrt{2}}{8}\cos(\epsilon)+\frac{4-\sqrt{2}}{8},\\
p(1|100)_{AB}=-\frac{\sqrt{2}}{8}\cos(\epsilon)+\frac{4-\sqrt{2}}{8},\\
p(1|101)_{AB}=\frac{\sqrt{2}}{8}\cos(\epsilon)+\frac{4+\sqrt{2}}{8},\\
p(1|110)_{AB}=-\frac{\sqrt{2}}{8}\cos(\epsilon)+\frac{4-\sqrt{2}}{8},\\
p(1|111)_{AB}=-\frac{\sqrt{2}}{8}\cos(\epsilon)+\frac{4-\sqrt{2}}{8}.
\end{array}
\end{equation}
By applying the previous conditional probability distributions
$p(b=1|x,y)_{AB}$, the dimension witness value  \cite{DW3,li1,li2}
between Alice and Bob is given by

\begin{equation}
\begin{array}{lll}
W_{1AB}=p(1|000)_{AB}+p(1|001)_{AB}+p(1|010)_{AB}\\
~~~~~~~~~~-p(1|011)_{AB}-p(1|100)_{AB}+p(1|101)_{AB}\\
~~~~~~~~~~-p(1|110)_{AB}-p(1|111)_{AB}\\
~~~~~~~~=\sqrt{2}(\cos(\epsilon)+1).
\end{array}
\end{equation}

Based on the previous state preparation and measurement operator, we
can calculate the corresponding conditional probability distribution
$p(b=1|x,z)_{AC}$ as follows

\begin{equation}
\begin{array}{lll}
p(1|000)_{AC}=\frac{2+\sqrt{2}}{4}\sin^2(\epsilon),\\
p(1|001)_{AC}=\frac{2+\sqrt{2}}{4}\sin^2(\epsilon),\\
p(1|010)_{AC}=\frac{2+\sqrt{2}}{4}\sin^2(\epsilon),\\
p(1|011)_{AC}=\frac{2-\sqrt{2}}{4}\sin^2(\epsilon),\\
p(1|100)_{AC}=\frac{2-\sqrt{2}}{4}\sin^2(\epsilon),\\
p(1|101)_{AC}=\frac{2+\sqrt{2}}{4}\sin^2(\epsilon),\\
p(1|110)_{AC}=\frac{2-\sqrt{2}}{4}\sin^2(\epsilon),\\
p(1|111)_{AC}=\frac{2-\sqrt{2}}{4}\sin^2(\epsilon).
\end{array}
\end{equation}
By applying the previous conditional probability distributions
$p(b=1|x,y)_{AC}$, the dimension witness value between Alice and
Charlie is given by

\begin{equation}
\begin{array}{lll}
W_{1AC}=2\sqrt{2}\sin^2(\epsilon).
\end{array}
\end{equation}

To prove local randomness generation in Bob's side, we analyze the
dimension witness value $W_{1ABz}$ with different input weak
measurement parameters $z\in\{0,1\}$.

In the case of $z=0$, we can obtain the corresponding conditional
probability distributions $p(b=1|x,y)_{AB,z=0}$ as follows

\begin{equation}
\begin{array}{lll}
p(1|000)_{AB,z=0}=\frac{2+\sqrt{2}}{4},\\
p(1|001)_{AB,z=0}=\frac{2+\sqrt{2}\cos(\epsilon)}{4},\\
p(1|010)_{AB,z=0}=\frac{2+\sqrt{2}}{4},\\
p(1|011)_{AB,z=0}=\frac{2-\sqrt{2}\cos(\epsilon)}{4},\\
p(1|100)_{AB,z=0}=\frac{2-\sqrt{2}}{4},\\
p(1|101)_{AB,z=0}=\frac{2+\sqrt{2}\cos(\epsilon)}{4},\\
p(1|110)_{AB,z=0}=\frac{2-\sqrt{2}}{4},\\
p(1|111)_{AB,z=0}=\frac{2-\sqrt{2}\cos(\epsilon)}{4}.
\end{array}
\end{equation}
By applying the previous conditional probability distributions
$p(b=1|x,y)_{AB,z=0}$, the dimension witness value between Alice and
Bob is given by

\begin{equation}
\begin{array}{lll}
W_{1AB,z=0}=\sqrt{2}\cos(\epsilon)+\sqrt{2}.
\end{array}
\end{equation}

In the case of $z=1$, we obtain the corresponding conditional
probability distribution $p(b=1|x,y)_{AB,z=1}$ as follows

\begin{equation}
\begin{array}{lll}
p(1|000)_{AB,z=1}=\frac{2+\sqrt{2}\cos(\epsilon)}{4},\\
p(1|001)_{AB,z=1}=\frac{2+\sqrt{2}}{4},\\
p(1|010)_{AB,z=1}=\frac{2+\sqrt{2}\cos(\epsilon)}{4},\\
p(1|011)_{AB,z=1}=\frac{2-\sqrt{2}}{4},\\
p(1|100)_{AB,z=1}=\frac{2-\sqrt{2}\cos(\epsilon)}{4},\\
p(1|101)_{AB,z=1}=\frac{2+\sqrt{2}}{4},\\
p(1|110)_{AB,z=1}=\frac{2-\sqrt{2}\cos(\epsilon)}{4},\\
p(1|111)_{AB,z=1}=\frac{2-\sqrt{2}}{4}.
\end{array}
\end{equation}
By applying the previous conditional probability distributions
$p(b=1|x,y)_{AB,z=1}$, the dimension witness value between Alice and
Bob is given by

\begin{equation}
\begin{array}{lll}
W_{1AB,z=1}=\sqrt{2}\cos(\epsilon)+\sqrt{2}.
\end{array}
\end{equation}

\subsection{ Quantum dimension witness based on determinant value }

 To
prove the double classical dimension witness violation under Eq.
(16), we propose the following density matrix preparation in Alice's
side
\begin{equation}
\begin{array}{lll}
\rho_{00}=(0,0,1),\\
\rho_{01}=(0,0,-1),\\
\rho_{10}=(1,0,0),\\
\rho_{11}=(-1,0,0).
\end{array}
\end{equation}
The measurement operator in Bob's side is given by
\begin{equation}
\begin{array}{lll}
\nu_{0}=(0,0,1),\\
\nu_{1}=(1,0,0).
\end{array}
\end{equation}
The measurement operator in Charlie's side is given by
\begin{equation}
\begin{array}{lll}
t=(1,0,0).
\end{array}
\end{equation}
The rotation operator in Charlie's side is given by
\begin{equation}
\begin{array}{lll}
\omega_{0}=(0,0,1),\\
\omega_{1}=(1,0,0).
\end{array}
\end{equation}
Note that the state preparation and measurement operator are the
same as the Bennett and Brassard 1984 (BB84) quantum key
distribution protocol \cite{BB84}, but the weak measurement model
may disturb Alice's quantum state with the different weak
measurement parameter $\epsilon$, and this disturbance can be
detected by the quantum bit error rate.

Based on the previous state preparation and measurement, we can
calculate the corresponding conditional probability distribution
$p(b=1|x,y)_{AB}$ as follows

\begin{equation}
\begin{array}{lll}
p(1|000)_{AB}=\frac{1}{4}\cos(\epsilon)+\frac{3}{4},\\
p(1|001)_{AB}=\frac{1}{2},\\
p(1|010)_{AB}=-\frac{1}{4}\cos(\epsilon)+\frac{1}{4},\\
p(1|011)_{AB}=\frac{1}{2},\\
p(1|100)_{AB}=\frac{1}{2},\\
p(1|101)_{AB}=\frac{1}{4}\cos(\epsilon)+\frac{3}{4},\\
p(1|110)_{AB}=\frac{1}{2},\\
p(1|111)_{AB}=-\frac{1}{4}\cos(\epsilon)+\frac{1}{4}.
\end{array}
\end{equation}
By applying the previous conditional probability distributions
$p(b=1|x,y)_{AB}$, the dimension witness value
\cite{brunner2,experiment1} between Alice and Bob is given by

\begin{equation}
\begin{array}{lll}
W_{2AB}\\
=\left|\begin{array}{ccc}
p(1|000)_{AB}-p(1|010)_{AB}&p(1|100)_{AB}-p(1|110)_{AB}\\
p(1|001)_{AB}-p(1|011)_{AB}&p(1|101)_{AB}-p(1|111)_{AB}\end{array}\right|\\
=(\frac{1}{2}\cos(\epsilon)+\frac{1}{2})^2.
\end{array}
\end{equation}

Based on the previous state preparation and measurement, we can also
calculate the corresponding conditional probability distribution
$p(b=1|x,z)_{AC}$ as follows

\begin{equation}
\begin{array}{lll}
p(1|000)_{AC}=\sin^2(\epsilon),\\
p(1|001)_{AC}=\frac{1}{2}\sin^2(\epsilon),\\
p(1|010)_{AC}=0,\\
p(1|011)_{AC}=\frac{1}{2}\sin^2(\epsilon),\\
p(1|100)_{AC}=\frac{1}{2}\sin^2(\epsilon),\\
p(1|101)_{AC}=\sin^2(\epsilon),\\
p(1|110)_{AC}=\frac{1}{2}\sin^2(\epsilon),\\
p(1|111)_{AC}=0.
\end{array}
\end{equation}
By applying the previous conditional probability distributions
$p(b=1|x,y)_{AC}$, the dimension witness value between Alice and
Charlie is given by

\begin{equation}
\begin{array}{lll}
W_{2AC}=\sin^4(\epsilon).
\end{array}
\end{equation}

To prove local randomness generation in Bob's side, we will analyze
the dimension witness value $W_{2ABz}$ with different input weak
measurement parameters $z\in\{0,1\}$.

In the case of $z=0$, we can obtain the corresponding conditional
probability distribution $p(b=1|x,y)_{AB,z=0}$ as follows

\begin{equation}
\begin{array}{lll}
p(1|000)_{AB,z=0}=1,\\
p(1|001)_{AB,z=0}=\frac{1}{2},\\
p(1|010)_{AB,z=0}=0,\\
p(1|011)_{AB,z=0}=\frac{1}{2},\\
p(1|100)_{AB,z=0}=\frac{1}{2},\\
p(1|101)_{AB,z=0}=\frac{1+\cos(\epsilon)}{2},\\
p(1|110)_{AB,z=0}=\frac{1}{2},\\
p(1|111)_{AB,z=0}=\frac{1-\cos(\epsilon)}{2}.
\end{array}
\end{equation}

Based on the previous conditional probability distributions
$p(b=1|x,y)_{AB,z=0}$, the dimension witness value between Alice and
Bob is given by

\begin{equation}
\begin{array}{lll}
W_{2AB,z=0}=\cos(\epsilon).
\end{array}
\end{equation}

In the case of $z=1$, we can obtain the corresponding conditional
probability distribution $p(b=1|x,y)_{AB,z=1}$ as follows

\begin{equation}
\begin{array}{lll}
p(1|000)_{AB,z=1}=\frac{1+\cos(\epsilon)}{2},\\
p(1|001)_{AB,z=1}=\frac{1}{2},\\
p(1|010)_{AB,z=1}=\frac{1-\cos(\epsilon)}{2},\\
p(1|011)_{AB,z=1}=\frac{1}{2},\\
p(1|100)_{AB,z=1}=\frac{1}{2},\\
p(1|101)_{AB,z=1}=1,\\
p(1|110)_{AB,z=1}=\frac{1}{2},\\
p(1|111)_{AB,z=1}=0.
\end{array}
\end{equation}
By applying the previous conditional probability distributions
$p(b=1|x,y)_{AB,z=1}$, the dimension witness value between Alice and
Bob is given by

\begin{equation}
\begin{array}{lll}
W_{2AB,z=1}=\cos(\epsilon).
\end{array}
\end{equation}

\end{thebibliography}
\end{document}